\documentclass[onecolumn]{revtex4}

\usepackage{epsfig}

\begin{document}

\title{On the spectral functions of scalar mesons}
\author{Francesco Giacosa and Giuseppe Pagliara}

\begin{abstract}
In this work we study the spectral functions of scalar mesons in one- and
two-channel cases. When the propagators satisfy the K\"allen-Lehman
representation a normalized spectral function is obtained, allowing to take
into account finite-width effects in the evaluation of decay rates. In the
one-channel case, suitable to the light sigma and k mesons, the spectral
function can deviate consistently from a Breit-Wigner shape. In the
two-channel case with one subthreshold channel the evaluated spectral
function is well approximated by a Flatte' distribution; when applying the
study to the $a_0(980)$ and $f_0(980)$ mesons the tree-level forbidden KK
decay is analysed.
\end{abstract}









\address{Institut f\"ur Theoretische Physik,
 Universit\"at Frankfurt,
 Johann Wolfgang Goethe - Universit\"at,
 Max von Laue--Str. 1
 60438 Frankfurt, Germany}

\maketitle

\section{Introduction}

The scalar mesons below $2$ GeV are in center of debate since many years 
\cite{amslerrev,closerev,exotica,bugg}. More states than expected from the
quark-antiquark assignment are reported in Particle Data Group (PDG) \cite%
{pdg}, leading to the introduction of a scalar glueball \cite{close95},
tetraquark states \cite{jaffe} and mesonic molecules \cite{mesonicmol1}. In
particular, the scalar resonances below 1 GeV have appealing
characteristics, such as the reversed level ordering of masses, expected
from tetraquark states \cite{exotica,jaffe,maiani,achasov1,tq}. In turn,
this scenario implies that quarkonia lie between $1$ and $2$ GeV. A
complication in the analysis of scalar states is mixing: between 1-2 GeV a
quarkonia-glueball mixing in the isoscalar sector is considered, for
instance, in Refs. \cite{closekirk}. Mixing among tetraquark states below 1
GeV and quarkonia above 2 GeV is studied in Refs. \cite%
{shechter,fariborz,tqchiral}, where, however, the results do not coincide:
while a large mixing is found in Ref. \cite{fariborz}, a negligible mixing
is the outcome of \cite{tqchiral}. It should be stressed that different
interpretations of scalar state are possible: a nonet of scalar quarkonia is
settled below 1 GeV in Refs. \cite{scadron} in agreement with the linear
sigma model and the Nambu Jona-Lasinio (NJL) model, while in Ref. \cite%
{minkowski} a broad glueball, to be identified with $f_{0}(600)$, is
proposed. We refer to Refs. \cite{amslerrev,closerev,exotica,achasov1} for a
discussion of arguments in favour and/or against the outlined assignments.

Studies on scalar mesons have been extensively performed by using chiral
perturbation theory \cite{colangelo}, where a scalar resonance at about $440$
MeV is inferred out of pion-pion scattering. A full nonet of molecular-like
scalar states is generated in the unitarized chiral perturbation theory of
Ref. \cite{oller}. In particular, Pelaez \cite{pelaez} studied the large-$%
N_{c}$ dependence of the light scalar resonances finding that they do not
scale as quarkonia but in agreement with a molecular or tetraquark composition 
(see, however, also the discussions in Refs. \cite{ishida}).

In the present paper we concentrate on an important aspect of light scalar
resonances, namely the form of their spectral functions, in a simple
theoretical context. In this study, relevant for both quarkonium or tetraquark assignment of light scalars
\cite{achasovprop}, effects arising from loops of pseudoscalar
mesons are considered: this leads to parametrizations of spectral functions beyond the
(usually employed) Breit-Wigner and Flatt\`{e} distributions and allow to
include finite-width effects in the evaluation of decay rates.
In particular, we consider the following
physical scenarios: (i) the case of a broad scalar resonance, strongly
coupled to one decay channel, such as the $\sigma \equiv f_{0}(600)$ in the
pion-pion decay mode, for which the spectral function can deviate
substantially form the Breit-Wigner form; (ii) the case of two channels, one
of which is sub-threshold for the mass and thus forbidden at tree-level,
as the $\overline{K}K$ decay mode for the resonances $f_{0}(980)$ and $%
a_{0}(980).$ In the latter case a comparison with the usually employed Flatt%
\`{e} distribution is performed.

The key-quantity of the discussion is the propagator of scalar resonances
dressed by mesonic loops in one or more channels. When the K\"{a}%
llen-Lehmann representation is satisfied, as verified at one-loop level in
the case of light scalar mesons for large ranges of parameters \cite%
{achasovprop}, the spectral function (proportional to the imaginary part of
the propagator) is correctly normalized and is interpreted as a `mass
distribution' for the scalar state. A general definition of the decay of a
scalar state, involving the obtained mass distribution, is then possible. In
this way one takes into account in a consistent fashion finite-width effects
for the decay, hence allowing to study deviations form the usually employed
tree-level formula for decay rates. Furthermore, the fulfillment of the K%
\"{a}llen-Lehmann representation offers a criterion to delimit the validity
of our one-loop study: as soon as violations appear (generally for large
coupling constants) the obtained spectral functions are no longer usable.

We regularize the mesonic loops by means of a cutoff function which in turn
is equivalent to a nonlocal interaction Lagrangian. In a phenomenological
perspective it is reasonable that the mesonic states in the loop cannot have
indefinitely high virtual momenta which are naturally limited due to the
finite range of the meson-meson interaction. We also show that the
dependence on the chosen cutoff function and on the specific value of the
cutoff is mild.

In order to render the paper easily understandable and self-contained we
start in Section II with one-channel case by recalling the basic definitions
and properties, then we apply the study to the scalar sigma $\sigma $ and
kaon $k$ resonances: the corresponding spectral function shows consistent
deviations from the usual Breit-Wigner one. In Section III we turn to the
two-channel case, with particular attention to the resonances $a_{0}(980)$
and $f_{0}(980)$, their decay rates and spectral functions in comparison
with the Flatt\`{e} distribution \cite{flatteorig,flatte}. Implications of
the results in view of a nonet of tetraquark states below 1 GeV is
discussed. In section IV we drive our conclusions, emphasizing as in Ref. 
\cite{achasovprop} that the use of propagators fulfilling the K\"{a}%
llen-Lehmann representation, which implies normalized distributions and a
correct definition of decay rates, should be preferable both in theoretical
and experimental works.

\section{Scalar spectral function: one-channel case}

\subsection{ Definitions and properties}

We consider the scalar fields $S$ and $\varphi $ described by the Lagrangian%
\begin{equation}
\mathcal{L}_{S}^{1}=\frac{1}{2}(\partial _{\mu }S)^{2}-\frac{1}{2}%
M_{0}^{2}S^{2}+\frac{1}{2}(\partial _{\mu }\varphi )^{2}-\frac{1}{2}%
m^{2}\varphi ^{2}+gS\varphi ^{2}.  \label{toy1ch}
\end{equation}%
In the limit $g=0$ the propagator of the field $S$ reads%
\begin{equation}
\Delta _{S}(p)=\frac{1}{p^{2}-M_{0}^{2}+i\varepsilon },\text{ }g=0.
\label{prop1}
\end{equation}%
We intend to study the modification to $\Delta _{S}(p)$ when $g\neq 0,$
which arises by considering the loop-diagram of Fig. 1 and how this contribution
affects the decay mechanism $S\rightarrow \varphi \varphi .$ We recall that at tree-level
the decay width reads
\begin{equation}
\Gamma _{S\varphi \varphi }^{\text{t-l}}(M_{0})=\frac{p_{S\varphi \varphi }}{%
8\pi M_{0}^{2}}[g_{S\varphi \varphi }]^{2}\theta (M_{0}-2m)\hspace{0.2cm} , \hspace{0.2cm}
p_{S\varphi \varphi }=\sqrt{\frac{M_{0}^{2}}{4}-m^{2}},\text{ }g_{S\varphi
\varphi }=\sqrt{2}g. 
\label{tl1}
\end{equation}%
where $\theta (x)$ is the step function and%


(The factor $\sqrt{2}$ in the amplitude $g_{S\rightarrow \varphi \varphi }$
takes into account that the final state consists of two identical
particles.) In general, under the symbol $p_{SAB}$ the expression 
\begin{equation}
p_{SAB}=\frac{1}{2M_{S}}\sqrt{%
M_{S}^{4}+(M_{A}^{2}-M_{B}^{2})^{2}-2(M_{A}^{2}+M_{B}^{2})M_{S}^{2}}
\end{equation}%
i.e. the momentum of the outgoing particle(s), is understood.

At tree-level the particle $S$ is treated as stable. However, the very fact
that the decay $\Gamma _{S\varphi \varphi }^{\text{t-l}}\neq 0$ for $%
M_{0}>2m $ means that $S$ is not stable and cannot be considered as an
asymptotic state of the Lagrangian $\mathcal{L}_{S}^{1}.$ The tree-level
expression $\Gamma _{S\varphi \varphi }^{\text{t-l}}$ is only valid in the
limit $g\rightarrow 0.$ The evaluation of the loop of Fig. 1 offers a way to
define and interpret the decay $S\rightarrow \varphi \varphi $ as we
describe in the following. The modified propagator of $S$ is obtained by
(re)summing the loop-diagrams of Fig. 1:%
\begin{equation}
\Delta _{S}(p^{2})=\frac{1}{p^{2}-M_{0}^{2}+g_{_{S\varphi \varphi
}}^{2}\Sigma (p^{2})+i\varepsilon }
\end{equation}%
where the self-energy $\Sigma (p^{2})$ reads: 
\begin{equation}
\Sigma (p^{2})=-i\int \frac{d^{4}q}{(2\pi )^{4}}\frac{1}{\left[
(q+p/2)^{2}-m^{2}+i\varepsilon \right] \left[ (q-p/2)^{2}-m^{2}+i\varepsilon %
\right] }.
\end{equation}%
The integral defining $\Sigma (p^{2})$ is, as known, logarithmic divergent.
Our intention is to consider the Lagrangian $\mathcal{L}_{S}^{1}$ as an
effective low-energy description of the fields $S$ and $\varphi ,$ and not
as a fundamental theory valid up to indefinitely high mass scales. We do not
apply the renormalization scheme to $\mathcal{L}_{S}^{1}$ but we introduce a
regularization function $f_{\Lambda }(q)$ which depends on a cut-off
scale $\Lambda $ for the large momenta. The self-energy $\Sigma (p^{2})$ is then modified to:%
\begin{equation}
\Sigma (p^{2})=-i\int \frac{d^{4}q}{(2\pi )^{4}}\frac{f_{\Lambda }^{2}(q^{o},%
\overrightarrow{q})}{\left[ (q+p/2)^{2}-m^{2}+i\varepsilon \right] \left[
(q-p/2)^{2}-m^{2}+i\varepsilon \right] }  \label{loopf}
\end{equation}%
When choosing $f_{\Lambda }(q)=f_{\Lambda }(q^{2})$ the covariance of the
theory is preserved, otherwise is lost. Indeed, in many calculations related
to mesonic loops a regularization of the kind $f_{\Lambda }(q)=f_{\Lambda }(%
\overrightarrow{q}^{2})$ is chosen, which leads to simple expressions for
the self-energy contribution but breaks covariance explicitly, and thus is
strictly valid only in the rest frame of the decaying particle. In
particular, the 3d-cutoff $f_{\Lambda }(\overrightarrow{q}^{2})=\theta
(\Lambda ^{2}-\overrightarrow{q}^{2})$ is often used. We refer to Appendix A
for a closer analysis of the self-energy $\Sigma (p^{2})$, where the case of
unequal masses circulating in the loop is also presented. The interaction
strength among light mesons is suppressed for distances larger than $l\sim
0.5$-$1$ fm: in this particular physical example it is then natural to
implement a cutoff $\Lambda \sim 1/l,$ which varies between $1$ and $2$ GeV .

\begin{figure}[tbp]
\begin{center}
\includegraphics[scale=0.5]{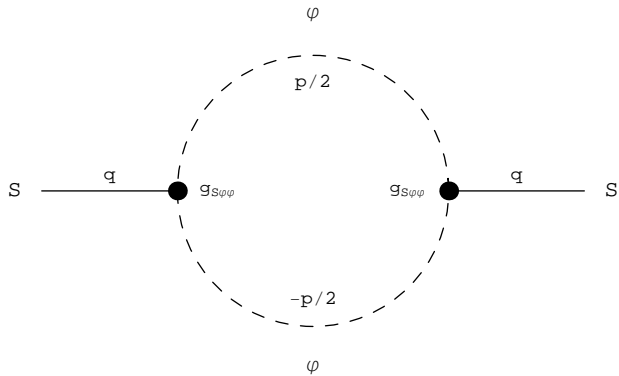}
\end{center}
\caption{Mesonic loop}
\label{loop}
\end{figure}

The cutoff-function $f_{\Lambda }(q)$ is not present in the Lagrangian $%
\mathcal{L}_{S}^{1}$ of Eq. (\ref{toy1ch}). In this sense the Lagrangian is
incomplete because it does not specify how to cut the high momenta. One can
take into account $f_{\Lambda }(q)$ already at the Lagrangian level by
rendering the interaction term nonlocal:%
\begin{equation}
\left( \mathcal{L}_{S}^{1}\right) _{int}=gS\varphi ^{2}\rightarrow gS\int
d^{4}y\varphi (x+y/2)\varphi (x-y/2)\Phi (y).  \label{deloc}
\end{equation}%
The Feynman rule for the 3-leg vertex is modified as:%
\begin{equation}
ig\rightarrow igf_{\Lambda }(\frac{q_{1}-q_{2}}{2}),\text{ }f_{\Lambda
}(q)=\int d^{4}y\Phi (y)e^{-iyq}  \label{ft}
\end{equation}%
where $q_{1}$ and $q_{2}$ are the momenta of the two particles $\varphi .$
The function $f_{\Lambda }(q)$ enters directly into the expression of all
amplitudes. In particular, the self-energy contribution of Eq. (\ref{loopf})
is now obtained by application of the (modified) Feynman rules to the loop
diagram of Fig. 1. Indeed, the delocalization of the interaction term also
induces a change of the tree-level result for the decay, which becomes (for
the case $f_{\Lambda }(q)=f_{\Lambda }(\overrightarrow{q}^{2})$):%
\begin{equation}
\Gamma _{S\varphi \varphi }^{\text{t-l}}(M_{0})=\frac{p_{S\varphi \varphi }}{%
8\pi M_{0}^{2}}[g_{S\varphi \varphi }f_{\Lambda }(\overrightarrow{q}^{2}=
p_{S\varphi \varphi }^{2})]^{2}\theta (M_{0}-2m),\text{ }g_{S\varphi
\varphi }=\sqrt{2}g  \label{tlnl}
\end{equation}%

that is the function $f_{\Lambda }(\overrightarrow{q}^{2})$ is explicitly
present in the tree-level decay expression and can be interpreted as a

phenomenological form factor \footnote{
For a covariant vertex function $f_{\Lambda }(q)$ the decay-amplitude takes the form 
$[g_{S \phi \phi}f_{\Lambda}(q^0=0,{\bf q}^2=p_{S\varphi \varphi }^{2})]$}


If a step-function is used the local
tree-level expression of Eq. (\ref{tl1}) is recovered, provided that the
cutoff $\Lambda $ is large enough. In this work we use the following
cutoff-function 

\begin{equation}
f_{\Lambda }(q)=f_{\Lambda }(\overrightarrow{q}^{2})=\left( 1+%
\overrightarrow{q}^{2}/\Lambda ^{2}\right) ^{-1}.  \label{cutoff}
\end{equation}%
With this choice the Fourier transform $\Phi (y)$, see Eq. (\ref{ft}), takes
the form $\delta (y^{0})\exp \left[ -\left\vert \overrightarrow{y}%
\right\vert \Lambda \right] /\left\vert \overrightarrow{y}\right\vert ,$
thus decreasing rapidly for increasing distance of the two interacting mesons $%
\varphi .$ The interaction range $l$ is of the order $\Lambda ^{-1},$ as
discussed above based on general dimensional grounds. At each step of the
forthcoming study we employed also different forms of $f_{\Lambda }(q)$, 
finding that the dependence on the precise form of $%
f_{\Lambda }(q)$ affects only slightly the results. Notice that in Refs. 
\cite{torn} a similar equation to (\ref{tlnl}) (where $s=\overrightarrow{q}%
^{2}$ and $f_{\Lambda }=G(s)$ in the notation of \cite{torn}) represents the
starting point of the analysis. The function $G(s)$ in the above cited works
is taken to be a Gaussian, $\Lambda $ is of the order of 1 GeV. The present
approach shows the link between such form factor $f_{\Lambda }=G(s)$ and a
nonlocal Lagrangian. However, we will not concentrate as in \cite{torn} on
scattering amplitudes but on spectral functions and decay widths. At the
same time we do not relate the imaginary and real part of the propagator via
K\"{a}llen-Lehmann dispersion relation, but we evaluate them
independently and subsequently we check numerically if it satisfied, see details in the
following discussion. 

Let us now turn to the self-energy $\Sigma (p^{2}).$ A general property for $%
\Sigma (p^{2})$ follows from the optical theorem:%
\begin{equation}
I(x)=g_{S\phi \phi }^{2}Im[\Sigma (x=\sqrt{p^{2}})]=x\Gamma
_{S\phi \phi }^{\text{t-l}}(x).  \label{I(x)}
\end{equation}%

The imaginary part of the self-energy diagram is zero for $0<x<2m$ and
nonzero starting at threshold. The real part 
\begin{equation}
R(x)=g_{S\varphi \varphi }^{2}Re[\Sigma (x=\sqrt{p^{2}})]
\end{equation}%
is nonzero below and above threshold. In Fig. 2 the functions $R(x)$ and $%
I(x)$ are plotted using Eq. (\ref{cutoff}). A particular choice for the
parameters $m=0.5$ GeV and $\Lambda =1.5$ GeV (of the order of physical
cases studied later) is done. Anyway the plotted functions are qualitatively
similar for large ranges of parameters. As noticeable, $R(x)$ is continuous
but not derivable in $x=2m.$ It has a cusp at $x=2m$: the left derivative is 
$+\infty ,$ while the right derivative is finite and negative.

In terms of the two functions $R(x)$ and $I(x)$ the propagator of Eq. (\ref%
{prop1}) reads%
\begin{equation}
\Delta _{S}(x)=\frac{1}{x^{2}-M_{0}^{2}+R(x)+iI(x)+i\varepsilon }.
\end{equation}%
We define the (Breit-Wigner) mass $M$ for the scalar field $S$ as the
solution of the equation 
\begin{equation}
M^{2}-M_{0}^{2}+R(M)=0.  \label{polem}
\end{equation}%
When the function $R(M)$ is positive, which is usually the physical case
(Fig. 2), the dressed mass $M$ is smaller than the bare mass $M_{0},$
showing that the quantum fluctuations tend to lower it.

\begin{figure}[tbp]
\begin{center}
\includegraphics[scale=0.7]{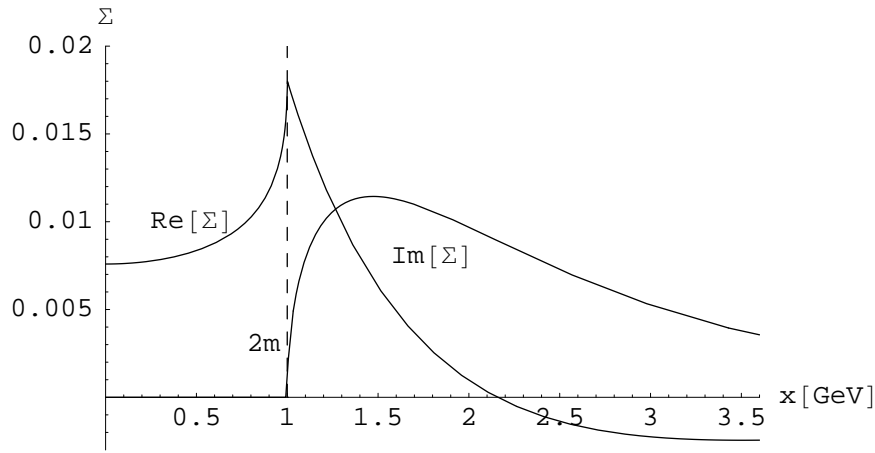}
\end{center}
\caption{Real and imaginary part of the mesonic loop for $m=0.5$ GeV and $%
\Lambda=1.5$ GeV. }
\label{re-im}
\end{figure}

We now turn to the spectral function $d_{S}(x)$ of the scalar field $S$
related to the imaginary part of the propagator as%
\begin{equation}
d_{S}(x)=\frac{2x}{\pi }\left\vert \lim_{\varepsilon \rightarrow 0}Im%
[\Delta _{S}(x)]\right\vert .  \label{dsdef}
\end{equation}%
In the limit $g\rightarrow 0$ we obtain the desired spectral function $%
d_{S}(x)=\delta (x-M_{0}).$ The normalization of $d_{S}(x)$ holds for each $g
$%
\begin{equation}
\int_{0}^{\infty }d_{S}(x)dx=1.  \label{norm}
\end{equation}%
The latter Eq. is a consequence of the K\"{a}llen-Lehmann representation%
\begin{equation}
\Delta _{S}(x)=\int_{0}^{\infty }dy \frac{2y}{\pi }\hspace{0.1cm}\frac{-Im[\Delta
_{S}(y)]}{y^{2}-x^{2}+i\varepsilon }  \label{kl}
\end{equation}%
when taking the limit $x\rightarrow \infty .$ Eqs. (\ref{norm}) and (\ref{kl}%
) hold in general for the full propagator. In our case we check numerically
the validity of the normalization condition (\ref{norm}) at one-loop level
of Fig. 1. We find that it is fulfilled to a high level of accuracy for
large ranges of parameters, see also the discussion in \cite{achasovprop}
and in the next subsection.

\begin{figure}[tbp]
\begin{center}
\includegraphics[scale=0.7]{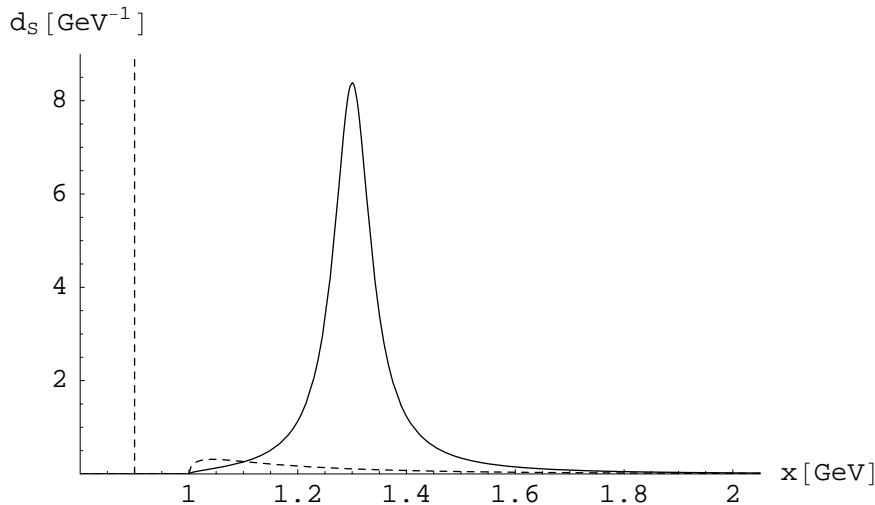}
\end{center}
\caption{Spectral functions in the cases $M<2m $ (dashed line) and $M>2m $
(continuous line). The coupling constant is $g_{S\protect\phi \protect\phi %
}=3$ GeV and two values of the masses are chosen: $M=0.9$ GeV and $M=1.3$
GeV, $m=0.5$ GeV.}
\label{spectralexample}
\end{figure}

Let us consider $d_{S}(x)$ in the two interesting cases $M<2m$ and $M>2m.$
If $M<2m$ Eq. (\ref{dsdef}) becomes: 
\begin{equation}
d_{S}(x)=Z\delta (x-M)\theta (2m-x)+\frac{2x}{\pi }\frac{I(x)}{\left(
x^{2}-M_{0}^{2}+R(x)\right) ^{2}+I(x)^{2}}  \label{dsbt}
\end{equation}%
where%
\begin{equation}
Z=\left( 1+\frac{1}{2M}\left( \frac{dR}{dx}\right) _{x=M}\right) ^{-1}.
\end{equation}%
When $M<2m$ the constant $Z$ is usually reabsorbed into the definition of
the wave function renormalization, hence recovering the free propagator
properly normalized as $\left( p^{2}-M^{2}+i\varepsilon \right) ^{-1}$,
corresponding to $d_{S}(x)=\delta (x-M)$ for $x<2m$ as in the free case $g=0$%
. Thus, we have still a stable particle with dressed mass $M$ instead of $%
M_{0}$. Notice that $0<Z<1$ because $R^{\prime }(M)$ is a positive number:
the quantity $\left( 1-Z\right) $ can be interpreted as the amount of
virtual clouds of $2\varphi $ contributing to the wave-function.

If $M>2m$ the spectral function reads%
\begin{equation}
d_{S}(x)=\frac{2x}{\pi }\frac{I(x)}{\left( x^{2}-M_{0}^{2}+R(x)\right)
^{2}+I(x)^{2}}.  \label{ds}
\end{equation}%
No delta-functions are present but typically a picked distribution $d_{S}(x)$
is obtained, corresponding to a physical resonance. The mass $M$ is not the
maximum of the $d_{S}(x)$ although in general very close to it. Consistent
deviations can appear when $M$ is close to threshold and for large coupling
constant, see next subsection for a more detailed discussion of this point.
Notice moreover that $d_{S}(x)$ is zero for $x<2m.$

We plot the typical behavior of the spectral function in both cases $M<2m$
and $M>2m$ in Fig.~\ref{spectralexample}. We used the values $M=0.9$ GeV and 
$M=1.3$ GeV corresponding to the the two cases below and above the
threshold, $m=0.5$ GeV as before and $g_{S\phi \phi }=3$ GeV. The value of $Z
$, for the sub-threshold case, is $\sim 0.9$, we have numerically verified
that the spectral functions are normalized in both cases.

When $M>2m$ the function $d_{S}(x)$ can be interpreted as the mass
distribution of the resonance, see also Appendix B for an intuitive
discussion about this point. We then define the decay rate for the process $%
S\rightarrow \varphi \varphi $ by implementing the distribution $%
d_{S}(x) $, and thus including finite width effects, as:%
\begin{equation}
\Gamma _{S\varphi \varphi }=\int_{0}^{\infty }dxd_{S}(x)\Gamma _{S\varphi
\varphi }^{\text{t-l}}(x).  \label{gendec}
\end{equation}%
This formula reduces to the tree-level amplitude $\Gamma _{S\varphi \varphi
}^{\text{t-l}}(M_{0})$ of Eq. (\ref{tlnl}) in the limit of small $g$: 
\begin{equation}
\Gamma _{S\varphi \varphi }^{\text{t-l}}(M_{0})\simeq \Gamma _{S\varphi
\varphi }\text{ for }g\rightarrow 0.
\end{equation}%
Notice that in this limit $M\rightarrow M_{0}$. However, even for finite $g,$
when $M\neq M_{0},$ the formula $\Gamma _{S\varphi \varphi }\simeq \Gamma
_{S\varphi \varphi }^{\text{t-l}}(M)$ offers a first approximation to the
decay width of the state as long as the distribution is picked, i.e. the
scalar state $S$ is not too broad.

The definition Eq. (\ref{gendec}) for the decay $S\rightarrow \varphi
\varphi $ is thus a generalization of the tree-level result of Eq. (\ref{tlnl}) and takes
automatically into account that the state $S$ has a finite width
parametrized by the mass distribution $d_{S}(x)$, which naturally arises by
considering the self-energy of the scalar propagator. Notice that the real
part of the propagator is necessary in order Eq. (\ref{norm}) to hold:
its neglection would spoil the correct normalization.

Evaluating the real and imaginary part at $x=M$ and neglecting their $x$%
-dependence, the distribution (\ref{ds}) is approximated by

\begin{equation}
d_{S}^{\text{bw}}(x)\simeq \frac{2M}{\pi }\frac{I(M)}{\left(
x^{2}-M^{2}\right) ^{2}+\left( I(M)\right) ^{2}}
\end{equation}%
which is the relativistic Breit-Wigner distribution for the resonance $S$,
usually employed in theoretical and experimental studies. However, the
distribution $d_{S}^{\text{bw}}(x)$ neglects the real part of the loop
diagram and consequently the normalization of Eq. (\ref{norm}) does not
hold, implying that $d_{S}^{\text{bw}}(x)$ has to be normalized by hand. At
the same time the mass $M$ does not coincide with the maximum of $d_{S}(x).$
Thus, we insist on that the usage of automatically normalized distribution
emerging from propagators fulfilling K\"{a}llen-Lehmann should be preferable.

\subsection{Application to the light scalar mesons $\protect\sigma $ and $k$}

An interesting example for the one-channel case is the decay of the scalar
meson $\sigma \equiv f_{0}(600)$. As reported by the PDG \cite{pdg},
experimental data are affected by large uncertainties both for the value of
the mass, $M_{\sigma }=0.4$-$1.2$ GeV, and the value of the Breit-Wigner
width, $\Gamma _{\sigma }=0.6$-$1$ GeV. The dominant channel, which we will
consider here, is the decay into two pions, for which $M>2m$ .

By applying the formulas of Section II.A we show in the left panel of Fig.~%
\ref{spect_sigma} the spectral functions $d_{\sigma }(x)$ of the $\sigma $%
-resonance for the two boundary cases of PDG, namely $M_{\sigma }=0.4$ GeV
and $M_{\sigma }=1.2$ GeV, respectively, for the coupling constant $%
g_{\sigma \pi \pi }=3$ GeV.

The spectral function assumes different shapes for different values of the
mass. While for $M_{\sigma }=1.2$, far from the threshold, the spectral
function has a regular `Breit-Wigner-like' form, in the case $M_{\sigma }=0.4$
GeV a distorted shape, with a narrow peak just above threshold, is visible%
\footnote{%
In the limit $M_{\sigma }\rightarrow 2m_{\pi }$ the spectral function $%
d_{S}(x)\sim 1/I(x)\rightarrow \infty $ for $x\rightarrow 2m_{\pi }$ due to
the threshold enhancement.}. The employed value of the coupling constant, $%
g_{\sigma \pi \pi }=3$ GeV, serves as illustration and actually corresponds
to a somewhat too narrow width. The increase of $g_{\sigma \pi \pi }$ leads,
however, outside the range of validity of the normalization of Eq. (\ref%
{norm}) at one-loop level, see below.

The description of the scalar kaonic resonance $k$ follows the same line 
\cite{vanbeveren}. As shown in the right panel of Fig.~\ref{spect_sigma} a
strong deviation from the Breit-Wigner form is obtained when $g_{k\pi K}=6$
GeV, corresponding to $\Gamma _{k}=0.38$ GeV, while a less distorted shape
is found for $g_{k\pi K}=3$ GeV, for which $\Gamma _{k}=0.11$ GeV.

\begin{figure}[tbp]
\begin{centering}
\hbox{\hskip-0.cm \epsfig{file=fig4.epsi,height=5cm}\hskip 0.3cm 
\epsfig{file=fig5.epsi,height=5cm}}
    \caption{Spectral functions of $\sigma$ and $k$.\\
Left panel: $\Lambda=1.5$ GeV, $g=3$ GeV, $M_{\sigma}=0.4$ GeV 
with a correspondent $\Gamma=0.242$ GeV and $M_{\sigma}=1.2$ GeV 
with a correspondent $\Gamma=0.113$ GeV.\\
Right panel: $M_{k}=0.8$ GeV, $\Lambda=1.5$ GeV, $g=3$ GeV 
with a correspondent $\Gamma=0.112$ GeV  and $g=6$ GeV 
with a correspondent $\Gamma=0.382$ GeV 
\label{spect_sigma}}
\end{centering}
\end{figure}

At this point a short discussion on the definition of the mass of an
unstable particle is needed: in the Breit-Wigner scheme, the mass of the
particle is the value corresponding to the maximum of the spectral function
and it is one of the parameter of the distribution (the second one is of
course the width). In our scheme, using the spectral functions coming from
the loop evaluation this is no more the case: the mass $M$ defined in Eq.~(%
\ref{polem}) is again a parameter of the distribution (together with the
coupling constant $g$) but it does not coincide with the maximum of the
distribution. While for $M_{\sigma }=1.2$ GeV (far from threshold) the
maximum of the spectral function occurs at $1.202$ GeV, thus only slightly
shifted from the mass, when $M_{\sigma }=0.4$ GeV the maximum of the
spectral function (apart form the threshold enhancement peak) occurs at
sizeable larger values respect to the mass, here $0.454$ GeV\footnote{%
Recent theoretical works \cite{colangelo} find a sigma mass at around $450$
MeV, thus not far from threshold in the lower side of the PDG data: this is
indeed the case of irregular form for the spectral function of this
resonance, for which care is needed.}. (Notice also that in the latter case
the bare mass $M_{0}$ is $0.614$ GeV, thus implying a strong influence of
the pion loop to the sigma mass, see also Appendix A for a comparison of
different `masses'). Indeed, although the mass $M$ being the zero of the
real part of the inverse propagator, see Eq. (\ref{polem}), is referred to
as Breit-Wigner mass \cite{torn}, the best fit to the full spectral function 
$d_{\sigma }(x)$ by using a Breit-Wigner form is obtained for a Breit-Wigner
mass $M_{BW}$ coinciding with the maximum of the distribution. It is also
remarkable that in some cases the spectral function has not a maximum, a
part from the threshold enhancement peak, as we can see for the $k$ meson in
the right panel of Fig.~\ref{spect_sigma} for $g_{k\pi K}=6$ GeV.

\begin{figure}[tbp]
\begin{center}
\includegraphics[scale=0.7]{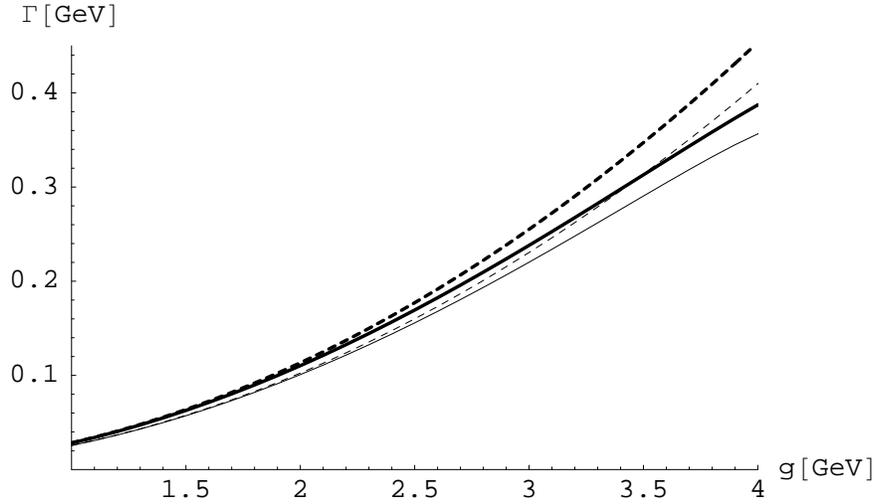}
\end{center}
\caption{The full (solid) and the tree-level (dashed) decay rates $\protect%
\sigma \rightarrow \protect\pi \protect\pi $ are shown as function of the
coupling constant. The cases $\Lambda =1$ and $\Lambda =2$ GeV correspond to
thin and thick lines respectively. }
\label{sigmadecay}
\end{figure}

We now study closer the decay process $\sigma \equiv f_{0}(600)\rightarrow
\pi \pi $ using Eq.~(\ref{gendec}) which implements the spectral function $%
d_{\sigma }(x).$ In Fig.~\ref{sigmadecay} we compare the full and the
tree-level decay rates for different values of the cut-off and of the
coupling constants (a mass $M_{\sigma }=600$ MeV is used).

As expected, there is not a strong dependence on the choice of the cut-off.
Moreover the results obtained with our formulae are well in agreement with
the tree-level results for small values of $g$ (since the spectral function
tends to a delta function). Non-negligible differences instead occur for the
larger values of $g$. This is due to a different analytic dependence of the
two corresponding formulae from the coupling constant: while the tree-level
expression depends quadratically on $g$, in the loop formula $g$ appears
also in the distribution $d_{S}(x)$.

The limit of the validity of the employed one-loop level analysis is an
important aspect which deserves further discussion. Namely, when the
coupling constant is too large and the mass is not far from threshold the
normalization condition of Eq. (\ref{norm}) is lost, see also the
corresponding discussion in \cite{achasovprop}. This fact means that higher
orders must be taken into account to satisfy the K\"{a}llen-Lehmann
representation and thus to recover the correct normalization of Eq. (\ref%
{norm}). At the same time the violation of the normalization is a valid
criterion to establish the limit of our study: for this reason in Fig. 5 we
stop the plot at $g_{\sigma \pi \pi }=4$ GeV (corresponding to $\Gamma
_{\sigma }\sim 400$ MeV), in fact larger values imply $\int_{0}^{\infty
}d_{S}(x)dx<1.$ At this point the full decay width (using Eq. (\ref{gendec}%
)) is already $\sim 60$ MeV smaller than the tree-level counterpart. A decay
width of about $400$ MeV is on the low side for the $\sigma $ (see \cite%
{ishida}). In Ref. \cite{colangelo} a width $150$ MeV larger is obtained. A
study beyond one-loop level would then be necessary to evaluate the spectral
function for larger coupling (i.e. larger width) and represents a possible
outlook of the present work. Surely the overestimation of the tree-level
formula keeps growing for increasing interaction strengths. Similar
considerations hold for the $k$ meson.

\section{Scalar spectral function: two-channel case}

\subsection{Definitions and properties}

We now consider two channels for the scalar resonance $S$ described by the
Lagrangian density ($m_{2}>m_{1}$): 
\begin{eqnarray}
\mathcal{L}_{S}^{2} &=&\frac{1}{2}(\partial _{\mu }S)^{2}-\frac{1}{2}%
M_{0}^{2}S^{2}+\frac{1}{2}(\partial _{\mu }\varphi _{1})^{2}-\frac{1}{2}%
m_{1}^{2}\varphi _{1}^{2}+  \nonumber \\
&&\frac{1}{2}(\partial _{\mu }\varphi _{2})^{2}-\frac{1}{2}m_{2}^{2}\varphi
_{2}^{2}+g_{1}S\varphi _{1}^{2}+g_{2}S\varphi _{2}^{2}.
\end{eqnarray}%
The processes $S\rightarrow \varphi _{1}\varphi _{1}$ and $S\rightarrow
\varphi _{2}\varphi _{2}$ correspond to the tree-level decay rates%
\begin{eqnarray}
\Gamma _{S\varphi _{1}\varphi _{1}}^{\text{t-l}}(M_{0}) &=&\frac{p_{S\varphi
_{1}\varphi _{1}}}{8\pi M_{0}^{2}}[g_{S\varphi _{1}\varphi _{1}}]^{2}\theta
(M_{0}-2m_{1}),\text{ }g_{S\varphi _{1}\varphi _{1}}=\sqrt{2}g_{1}, \\
\Gamma _{S\varphi _{2}\varphi _{2}}^{\text{t-l}}(M_{0}) &=&\frac{p_{S\varphi
_{2}\varphi _{2}}}{8\pi M_{0}^{2}}[g_{S\varphi _{2}\varphi _{2}}]^{2}\theta
(M_{0}-2m_{2}),\text{ }g_{S\varphi _{2}\varphi _{2}}=\sqrt{2}g_{2}\text{ .}
\end{eqnarray}%
The propagator is modified by loops of $\varphi _{1}$ and $\varphi _{2}$,
denoted as $\Sigma _{1}(p^{2})$ and $\Sigma _{2}(p^{2})$ and given by Eq. (%
\ref{loopf}) for $m=m_{1}$ and $m=m_{2}$ respectively. A delocalization of
the interaction, via a vertex-function $\Phi (y)$ and the corresponding
Fourier-transform $f_{\Lambda }(q)=\int d^{4}y\Phi (y)e^{-iyq},$ is then
introduced as in Eq. (\ref{deloc}) for both channels in order to regularize
the self-energy contributions. As a consequence, the tree-level results are
modified as%
\begin{eqnarray}
\Gamma _{S\varphi _{1}\varphi _{1}}^{\text{t-l}}(M_{0}) &=&\frac{p_{S\varphi
_{1}\varphi _{1}}}{8\pi M_{0}^{2}}[g_{S\varphi _{1}\varphi _{1}}f_{\Lambda }(%
\overrightarrow{q}^{2}=p_{S\varphi _{1}\varphi _{1}}^{2})]^{2}\theta
(M_{0}-2m_{1}) \\
\Gamma _{S\varphi _{2}\varphi _{2}}^{\text{t-l}}(M_{0}) &=&\frac{p_{S\varphi
_{2}\varphi _{2}}}{8\pi M_{0}^{2}}[g_{S\varphi _{2}\varphi _{2}}f_{\Lambda }(%
\overrightarrow{q}^{2}=p_{S\varphi _{2}\varphi _{2}}^{2})]^{2}\theta
(M_{0}-2m_{2}),
\end{eqnarray}%
and the propagator as

\begin{equation}
\Delta _{S}(x)=\frac{1}{x^{2}-M_{0}^{2}+R(x)+iI(x)+i\varepsilon }.
\end{equation}%
where%
\begin{equation}
R(x)=g_{S\varphi _{1}\varphi _{1}}^{2}Re[\Sigma _{1}(x=\sqrt{p^{2}}%
)]+g_{S\varphi _{2}\varphi _{2}}^{2}Re[\Sigma _{2}(x=\sqrt{p^{2}})]
\end{equation}%
and%
\begin{eqnarray}
I(x) &=&g_{S\varphi _{1}\varphi _{1}}^{2}Im[\Sigma _{1}(x=\sqrt{p^{2}}%
)]+g_{S\varphi _{2}\varphi _{2}}^{2}Im[\Sigma _{2}(x=\sqrt{p^{2}})] \\
&=&x\Gamma _{S\varphi _{1}\varphi _{1}}^{\text{t-l}}(x)+x\Gamma _{S\varphi
_{2}\varphi _{2}}^{\text{t-l}}(x).
\end{eqnarray}%
In the last equation the optical theorem has been used. The mass $M$ of the
state $S$ is given by $M^{2}-M_{0}^{2}+R(M)=0.$ Again, we have two cases:

(i) $M<2m_{1}$: the distribution $d_{S}(x)$ takes the form $Z\delta (x-M)$
for $x<2m_{1}.$ The discussion is similar to the one-channel case. Eq. (\ref%
{dsbt}) is still valid. At threshold $2m_{1}$ the continuum starts.

(ii) $M>2m_{1}$: as in Eq. (\ref{ds}) the distribution is 
\begin{equation}
d_{S}(x)=\frac{2x}{\pi }\frac{I(x)}{\left( x^{2}-M_{0}^{2}+R(x)\right)
^{2}+I(x)^{2}}.
\end{equation}%
It vanishes for $M<2m_{1}.$ At $x=2m_{2}$ the second channel opens.

In the case (ii) we have a resonant state. The decay rates into the two
channels $S\rightarrow \varphi _{1}\varphi _{1}$ and $S\rightarrow \varphi
_{2}\varphi _{2}$ are given by the integrals:%
\begin{equation}
\Gamma _{S\varphi _{1}\varphi _{1}}=\int_{0}^{\infty }dxd_{S}(x)\Gamma
_{S\varphi _{1}\varphi _{1}}^{\text{t-l}}(x),\text{ }\Gamma _{S\varphi
_{2}\varphi _{2}}=\int_{0}^{\infty }dxd_{S}(x)\Gamma _{S\varphi _{2}\varphi
_{2}}^{\text{t-l}}(x).  \label{dec2c}
\end{equation}

A particularly interesting case takes place when $2m_{1}<M<2m_{2}$: while
the tree-level result for $S\rightarrow \varphi _{2}\varphi _{2}$ vanishes,
we find that $\Gamma _{S\varphi _{2}\varphi _{2}}$ is not zero. In this case
the tree-level approximation is absolutely not applicable: the particle $S$
does decay in virtue of the high-mass tail of its distribution. A physical
example is well-known: the resonances $f_{0}(980)$ and $a_{0}(980)$ have a
non-zero decay rate into $\overline{K}K,$ although their masses are
below the threshold $2m_{K}.$ Clearly, a sizable decay rate $\Gamma
_{S\varphi _{2}\varphi _{2}}$ is obtained only when $M$ is close to
threshold. A generalization to the present definitions to $N$ channels is
straightforward \cite{achasovprop}.

When applying the decay formulas (\ref{dec2c}) it is however important to
verify numerically that the normalization of the distribution $d_{S}(x)$
holds: in fact, as discussed in Section II.B, only in this case the
formalism is self-consistent.

\subsection{Application to $a_{0}(980)$ and $f_{0}(980)$}

In this subsection we study the spectral functions of the scalar mesons $%
a_{0}\equiv a_{0}(980)$ and $f_{0}\equiv f_{0}(980)$, whose masses are $%
M_{a_{0}}=984.7\pm 1.2$ MeV and $M_{f_{0}}=980\pm 10$ MeV \cite{pdg}. For
both resonances two decays have been observed: $a_{0}\rightarrow \pi \eta $, $%
a_{0}\rightarrow K\overline{K}$ and $f_{0}\rightarrow \pi \pi $, $%
f_{0}\rightarrow K\overline{K}.$ Notice that both masses are below threshold
of kaon-antikaon production, $M_{a_{0},f_{0}}<2M_{K}=987.3$ MeV, thus the
decay of both resonances in $K\overline{K}$ vanishes at tree-level, while
experimentally it was seen for both $a_{0}$ and $f_{0}$ states.

\begin{figure}[tbp]
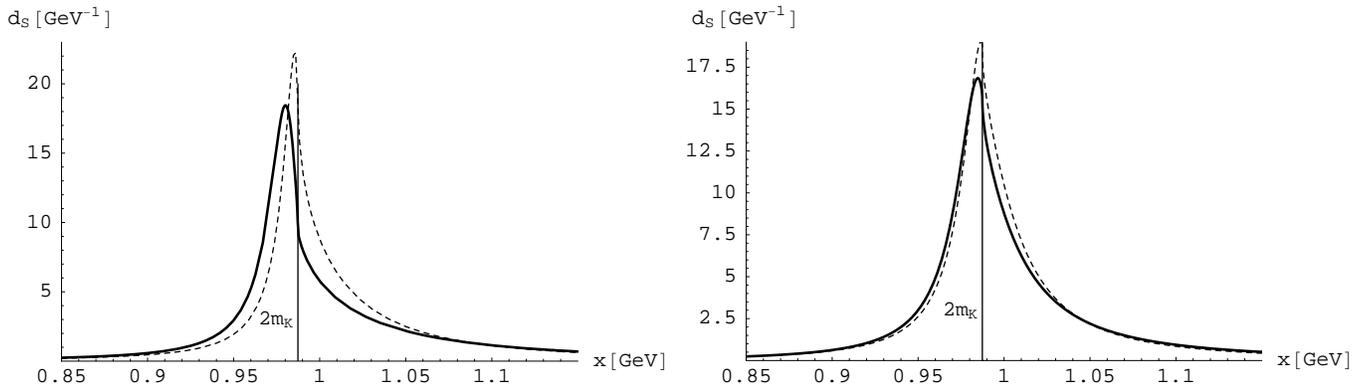

\begin{centering}
\hbox{\hskip-0.cm \epsfig{file=fig7.epsi,height=5cm}\hskip 0.3cm 
\epsfig{file=fig8.epsi,height=5cm}}
\caption{Left panel: spectral function of $f_0$ within our formalism (solid line) and 
using the Flatte' distribution (dashed line). Parameters are: $\Lambda=1.5$ GeV,
$g_{f_{0}K\overline{K}}=3$ GeV with a correspondent total decay rate 
$\Gamma=0.058$ GeV.\\
Right panel: spectral function of $a_0$ within our formalism (solid line) and 
using the Flatte' distribution (dashed line). Parameters are: $\Lambda=1.5$ GeV,
$g_{f_{0}K\overline{K}}=3$ GeV, with a correspondent total decay rate 
$\Gamma=0.048$ GeV. \label{distribuzioni-a0f0}}
\end{centering}
\end{figure}

For definiteness we use the following ratios obtained in the experimental
analysis of \cite{bugg}

\begin{equation}
\frac{g_{f_{0}K\overline{K}}^{2}}{{g_{f_{0}\pi \pi }^{2}}}=4.21\pm 0.46,\hspace{0.2cm}%
\frac{g_{f_{0}K\overline{K}}^{2}}{{g_{a_{0}K\overline{K}}^{2}}}=2.15\pm 0.40,\hspace{0.2cm}
\frac{g_{a_{0}\pi \eta }^{2}}{{g_{a_{0}K\overline{K}}^{2}}}=0.75\pm 0.11
\label{bugg}
\end{equation}%
therefore leaving us with only one free parameter, chosen to be $g_{f_{0}K%
\overline{K}}.$ Although experimental uncertainties are still large, the
results of Eq. (\ref{bugg}) are qualitative similar to various studies, see 
\cite{flatte} and Refs. therein, pointing to a large $K\overline{K}$
coupling for both resonances with a particular enhancement for $f_{0}$ (see 
\cite{bugg,achasov1,tqchiral,tq} and Refs. therein for spectroscopic
interpretations).

For the typical value $g_{f_{0}K\overline{K}}=3$ GeV we report in Fig.~\ref%
{distribuzioni-a0f0} the spectral functions of $a_{0}$ and $f_{0.}$ There is
a large probability, $\sim 50\%$, in both cases, that these two mesons have
a mass larger than the threshold of production $2m_{K}$ and therefore the
tree-level forbidden decay occurs. In the same figure we compare our
distribution with the Flatt\`{e} one \cite{flatteorig,flatte}, which is
usually employed for the $a_{0}$ and $f_{0}$ mesons. At variance from our
distribution, Flatt\`{e} distribution must be normalized by hand. The two
distributions are quite similar, only for the $f_{0}$ the values of the mass
corresponding to the maximum of the distributions are slightly different.
This is due to the strong coupling of $f_{0}$ to kaons and, as already
argued by Achasov \cite{achasovprop}, the meson loop distributions coincide
with the Flatt\`{e} ones only in the limit of weak coupling.

\begin{figure}[tbp]
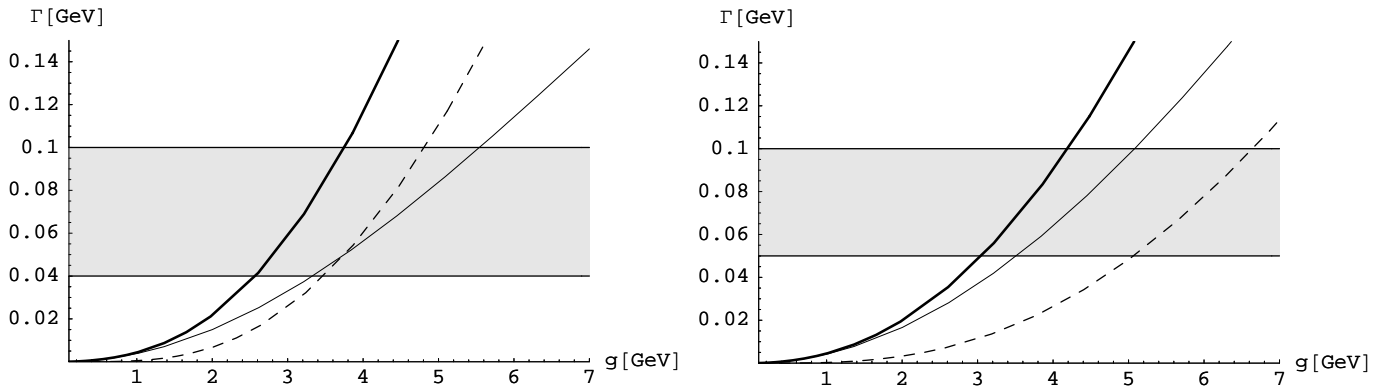

\begin{centering}
\hbox{\hskip-0.cm \epsfig{file=fig9.epsi,height=5cm}\hskip 0.3cm 
\epsfig{file=fig10.epsi,height=5cm}}
    \caption{Left panel: decay rates of $f_0$ as functions of the
coupling constant $g$. The thin solid line corresponds to the decay into two
pions, the dashed line to the decay into two kaons and the thick solid line
is the sum of the two decay rates.\\
Right panel: decay rates of $a_0$  as functions of the
coupling constant $g$. The thin solid line corresponds to the decay into pion
and eta, the dashed line to the decay into two kaons and the thick solid
line is the sum of the two decay rates.
\label{decaya0f0} }
\end{centering}
\end{figure}

In Fig.~\ref{decaya0f0} we show the decay rates $f_{0}\rightarrow \pi \pi $, 
$f_{0}\rightarrow K\overline{K}$ and $a_{0}\rightarrow \pi \eta $, $%
a_{0}\rightarrow K\overline{K}$ as function of $g_{f_{0}K\overline{K}}$. The
dashed areas in both plots correspond to the total decay rate of $f_{0}$ and 
$a_{0}$ as indicated by the PDG (notice, however, that in a note in PDG it
is specified that the real width could be larger). To get agreement between
our theoretical total decay rates and the measured ones, $g_{f_{0}K\overline{%
K}}$ has to lye between $3$-$4$ GeV. The outcoming branching ratio $\Gamma
_{a_{0}K\overline{K}}/\Gamma _{a_{0}\pi \eta }$ is $\sim 0.3$-$0.4,$ larger
than the PDG average $0.183\pm 0.024$, while the obtained ratio $%
\Gamma _{f_{0}\pi \pi }/(\Gamma _{f_{0}\pi \pi }+\Gamma _{f_{0}K\overline{K}%
})$ is $\sim 0.48$-$0.56$ in qualitative agreement with the (not bold)
results listed in PDG. Notice furthermore that the $f_{0}$ mesons turns out
to have typically a larger width than $a_{0}$ meson in agreement with Ref. 
\cite{flatte}.

We finally comment on a possible tetraquark unified interpretation of the
light scalar mesons as presented in Ref. \cite{tq}. A too small decay
constant $g_{f_{0}K\overline{K}}$ would also imply a by far too narrow $%
\sigma $ and $k$ mesons (related by Clebsh-Gordon coefficients \cite{tq}),
thus against a tetraquark nonet. On the contrary, $g_{f_{0}K\overline{K}}$
between $3$-$4$ GeV is in agreement with a tetraquark nonet below $1$ GeV,
although problems, such as a too narrow $k$, persist, see discussions in 
\cite{tq,tqchiral}. Such a strong coupling in the $K\overline{K}$-channel
implies that virtual cloud of kaon-antikaon pairs plays an important role,
in particular for the $f_{0}$ resonance. A heuristic indicator of the
mesonic cloud can be given by the quantity $Z_{\overline{K}K}=\left( 1+\frac{%
1}{2M_{S}}\left( \frac{dR_{\overline{SK}K}}{dx}\right) _{x=M_{S}}\right)
^{-1}$ where $R_{S\overline{K}K}=g_{S\overline{K}K}^{2}Re[\Sigma
_{KK}(x)]$ (with $S=f_{0},a_{0}$) refers to the kaonic loop only. As
discussed in Section II in the subthreshold case (which applies to the
kaonic channel here) the quantity $\left( 1-Z\right) $ varies between $0$
and $1$ and measures the mesonic cloud dressing of the original bare
resonance $S.$ In the $f_{0}$ case, by using Eq. (\ref{bugg}) together with $%
g_{f_{0}K\overline{K}}=3$ GeV, one finds $\left( 1-Z_{\overline{K}K}\right)
=0.38,$ hence implying a 38\% of kaonic cloud. This number increases for
increasing coupling strength $g_{f_{0}K\overline{K}}.$ This discussion
confirms the interpretation put forward in Ref. \cite{closerev}, where the
light scalar mesons posses a tetraquark core but are dressed by kaonic
clouds.

\section{Summary and conclusions}

In this work we studied the spectral functions of scalar mesons in one- and
two-channel cases suitable for the description of light scalar mesons below
1 GeV. We have computed, by using Lagrangians with non-derivative couplings,
the propagators of scalar mesons at one-loop level. They satisfy for large
ranges of parameters the K\"{a}llen-Lehmann representation, therefore
implying normalized spectral functions. In this way a correct definition of
decay amplitudes, weighted over the spectral function, is formulated: the
finite-width effects are automatically taken into account. The resulting
decay rates are smaller than the tree-level ones with increasing mismatch
for increasing interaction strength. On the other hand, a sub-threshold
tree-level forbidden decay, such as the $K\overline{K}$ mode for $a_{0}(980)$
and $f_{0}(980),$ becomes large.

The resulting spectral functions for the $\sigma $ and $k$ mesons may
deviate consistently from the Breit-Wigner form. The Flatt\`{e}
distribution, although it approximates to a good level of accuracy the $%
a_{0}(980)$ and $f_{0}(980)$ spectral functions, emerges as a small-coupling
limit of our more general spectral function.

As stressed by Achasov \cite{achasovprop} it is important to use
distributions satisfying K\"{a}llen-Lehmann representations in experimental
and theoretical studies. We thus believe that the use of distributions
obtained from quantum field theoretical models fulfilling the correct
normalization requirements can be helpful to correctly disentangle the
nature of the scalar states. Future studies with derivative couplings,
mixing effects and $\phi $-decays represent a possible interesting outlook.

{\bf Acknowledgements}     
G.P. acknowledges financial support from INFN.  

\appendix

\section{Loop contributions}

Here we report basic formulas for the loop diagram of Eq. (\ref{loopf})
drawn in Fig.1 for the vertex function $f_{\Lambda }(q)=f_{\Lambda }(%
\overrightarrow{q}^{2})$. By evaluating the residua one obtains the
one-dimensional integral%
\begin{equation}
\Sigma (x=p^{2})=\frac{1}{2\pi ^{2}}\int_{0}^{\infty }\mathrm{dw}\frac{%
w^{2}f_{\Lambda }^{2}(w)}{\sqrt{w^{2}+m^{2}}\left(
4(w^{2}+m^{2})-x^{2}\right) }  \label{lupo}
\end{equation}%
which can be easily evaluated numerically for each well-behaved $f_{\Lambda
}(w)$. We remind that within our conventions $f_{\Lambda }(0)=1$ and that $%
w=\left\vert \overrightarrow{k}\right\vert ;$ Eq. (\ref{lupo}) refers to a
3d-vertex function. In \cite{achasovprop} the form $f_{\Lambda }(w)=\theta
(\Lambda -w)$ is used and the limit $\Lambda \rightarrow \infty $ is taken.
As described in the text, we did not follow this procedure but we used
definite form(s) for the vertex function $f_{\Lambda }(w)$. As remarked in
the text, we performed the calculations also with different forms for $%
f_{\Lambda }(w)$ (different power form and exponential functions): the
precise form of the cutoff function does not affect the physical picture.

When the scalar state $S$ couples to two particles of masses $m_{1}$ and $%
m_{2}$ the loop contribution is modified as following:%
\begin{equation}
\Sigma (x)=\frac{1}{4\pi ^{2}}\int_{0}^{\infty }\mathrm{dw}\frac{w^{2}\left( 
\sqrt{w^{2}+m_{1}^{2}}+\sqrt{w^{2}+m_{2}^{2}}\right) f_{\Lambda }^{2}(w)}{%
\sqrt{w^{2}+m_{1}^{2}}\sqrt{w^{2}+m_{2}^{2}}\left( \left( \sqrt{%
w^{2}+m_{1}^{2}}+\sqrt{w^{2}+m_{2}^{2}}\right) ^{2}-x^{2}\right) }.
\end{equation}%
The choice $f_{\Lambda }(w)=\left( 1+w^{2}/\Lambda ^{2}\right) ^{-1}$ with $%
\Lambda =1$-$2$ GeV has been used in this work.

In relation to the mass definition of Section II.B we report and compare in
Table 1 the mass $M$ defined in Eq. (\ref{polem}), the bare mass $M_{0},$
the maximum $M_{\text{max}}$ of the distribution $d_{S\equiv \sigma }(x)$
and the average mass $\left\langle M\right\rangle =\int_{0}^{\infty }\mathrm{%
dx}xd_{S}(x)$. We use $m=m_{\pi },$ $g_{\sigma \pi \pi }=3$ GeV, $\Lambda
=1.5$ GeV.

\begin{center}
\textbf{Tab.1}: Comparison of `masses'

\begin{tabular}{|l|l|l|l|}
\hline
$M$ & $M_{0}$ & $M_{\text{max}}$ & $\left\langle M\right\rangle $ \\ \hline
0.4 & 0.61 & 0.45 & 0.54 \\ \hline
0.6 & 0.71 & 0.62 & 0.65 \\ \hline
0.8 & 0.86 & 0.81 & 0.82 \\ \hline
1 & 1.03 & 1.00 & 1.01 \\ \hline
\end{tabular}
\end{center}

As expected, the larger the mass, the smaller the differences among the
various mass-like quantities.

\section{Spectral function as `mass distribution': an intuitive discussion}

We present an intuitive argument for the correctness of interpretation of
the spectral function $d_{S}(x)$ as the `mass distribution' of the state $S$%
. To this end we introduce two scalar fields $A$ and $B,$ the first massless
and the second with $M_{B}>M_{S}$ and write down the interaction Lagrangian%
\begin{equation}
L=cBAS+gS\varphi ^{2}.
\end{equation}%
(for the following discussion the `delocalization' of Eq. (\ref{deloc}) is
not important). We suppose that the interaction strength $c$ is small enough
to allow a tree-level analysis for the decay of the state $B.$ The term $cBAS
$ generates the decay process $B\rightarrow AS,$ which reads (at tree-level)%
\begin{equation}
\Gamma _{BAS}^{\text{t-l}}(M_{B})=\frac{p_{BAS}}{8\pi M_{B}^{2}}[c]^{2}.
\label{BAS}
\end{equation}%
However, when $g\neq 0$ the state $S$ decays into $\varphi \varphi ,$ that
is the state $S$ is not an asymptotic state. Physically, we observe a
tree-body decay $B\rightarrow A\varphi \varphi ,$ whose decay-rate reads:%
\begin{equation}
\Gamma _{BA\varphi \varphi }^{\text{t-l}}(M_{B})=\int_{0}^{M_{B}}\Gamma
_{BAS}^{\text{t-l}}(M_{B})d_{S}(x)dx.  \label{3bd}
\end{equation}%
The tree-body decay is decomposed into two steps: $B\rightarrow AS$ and $%
S\rightarrow \varphi \varphi .$ The quantity $\Gamma _{BAS}^{\text{t-l}%
}(M_{B})$ represents the probability for $B\rightarrow AS$ (at a given mass $%
x$ for the state $S)$ and $d_{S}(x)dx$ is the corresponding weight, i.e. the
probability that the resonance $S$ has a mass between $x$ and $x+dx.$ In
this example $d_{S}(x)$ emerges naturally as a mass distribution, correctly
normalized, for the scalar state $S.$ Furthermore, notice that in virtue of
the limit $d_{S}(x)=\delta (M-M_{S})$ for $g\rightarrow 0$ one has 
\begin{equation}
\Gamma _{BA\varphi \varphi }^{\text{t-l}}(M_{B})=\Gamma _{BAS}^{\text{t-l}%
}(M_{B})\text{ for }g\rightarrow 0.
\end{equation}%
In fact, if $g$ is very small the state $S$ is long-living and the equation (%
\ref{BAS}) is recovered. The present analysis also shows that studies on the
tree-body decay of the $\phi $ meson can be consistent only if propagators
satisfying the K\"{a}llen-Lehmann representation are used.

\end{document}